\documentclass[10pt,letterpaper,twocolumn]{article} 

\usepackage{ol2}
\usepackage[draft]{hyperref}
\usepackage{amsmath}
\usepackage{amssymb}

\begin{document}

\twocolumn[ 

\title{Rb optical resonance inside a random porous medium}

\author{S. Villalba$^{1}$, H. Failache$^{1,*}$, A. Laliotis$^{2,1}$, L. Lenci$^{1}$, S. Barreiro$^{1}$ and A. Lezama$^{1}$}

\address{
$^1$Instituto de F\'{\i}sica, Facultad de Ingenier\'{\i}a,
Universidad de la Rep\'{u}blica,\\ J. Herrera y Reissig 565, 11300
Montevideo, Uruguay\\
$^2$Laboratoire de Physique des Lasers UMR 7538 du CNRS, Universit\'{e} Paris-13, F-93430, Villetaneuse, France\\
$^*$Corresponding author: heraclio@fing.edu.uy
}

\begin{abstract}We studied resonant laser interaction with Rb atoms confined to the
interstitial cavities of a random porous glass. Due to diffusive light propagation, the effect of atomic absorption on the light scattered by the sample is almost entirely compensated by atomic fluorescence at low atomic densities. For higher densities, radiation trapping increases the probability of non-radiative decay via atom-wall collisions. A simple connection of the fluorescence/absorption yield to the sample porosity is given.\end{abstract}


 ] 

\noindent Recent advances in the miniaturization and integration of nonlinear optical devices, have attracted considerable attention on atoms confined to micrometric or nanometric-size volumes such as hollow-core optical fibers \cite{Ghosh:2006, Slepkov:2010},
wave-guides \cite{Yang:2007}, thin cells \cite{Briaudeau:1999,
Dutier:2003b, Knappe:2004, Lenci:2009, Kubler:2010} or interstitial
cavities in nano-structures such as opals \cite{Ballin:2012}. Positronium atoms were created and observed in porous silica \cite{Cassidy:2011}. Here, the first spectroscopic study of alkali atoms confined to the interstitial cavities of a random porous medium is presented.

Porous media are strong light scatterers. Mean optical paths of several meters have been measured for $cm$-size porous ceramics used as compact gas cells for the detection of O$_2$ \cite{Svensson:2010, Svensson:2011}. Our sample is made of ground
glass with $50-100\ \mu m$ grain size. The scattering mean free path is $\sim 1\ mm$. Strong light scattering results in high randomization of the photon trajectories. As a consequence, the scattered light reaching an external detector tends to be proportional to the total light present in the medium no matter whether it originated from the laser or from atomic fluorescence. We call this the ``integrating-sphere effect" (ISE) by analogy with an extended
detector wrapping the full solid angle around the sample. A second influence of the medium porosity regards the laser-atom dynamics. Due to the smallness of the interstitial cavities, the laser-atom evolution has a significant probability to be interrupted by a collision with the pore wall \cite{Danos:1953,Ghosh:2006,Cassidy:2011,Svensson:2010,Svensson:2012}. Excited atoms may collide with the walls and relax their energy non-radiatively .

We have studied the optical resonance of Rb atoms contained inside porous glass.\\


In the linear regime, on resonance with an atomic transition, the steady state detected light power after the porous sample $I'$ is: $I^{\prime }=I\left( 1-\alpha +\beta \right)$
where $I$ is the off-resonance signal. Here $\alpha $ represents the relative light absorption and $\beta $ the light emission by excited atoms. The contributions of $\alpha $ and $\beta $ can be separated if one assumes that the absorption rapidly follows any change in the light intensity while the emission has a slower evolution due to the finite lifetime of the atomic excited state \cite{Yariv1989}.  If $I$ is switched at $t=0$ from $I_1$ to $I_2$, the transmitted light power is: $I^{\prime }\left( t>0\right) \simeq I_2\left( 1-\alpha +\beta \right)
+(I_1-I_2)\beta e^{-\frac{t}{\theta}}$. Here $\theta$ is the decay time of the slowly varying term. Thus, the transmitted light transient evolution allows the determination of $\theta$ and $\beta $, then $\alpha $ is obtained from the steady state transmission.

At high atomic densities, the characteristic decay time of the atomic fluorescence is affected by the onset of photon-trapping: emitted photons have a significant probability of being re-absorbed by other atoms. As the number of absorption-emission cycles grows, the decay time of the fluorescence is increased. A simple model of photon trapping can be sketched as follows \cite{supplemental}.

Let $p_{0}$ be the probability for a photon present in the atomic medium to be absorbed by an atom. For simplicity, $p_{0}$ is assumed independent of time, position or photon frequency. The single atom probability of emission at time $t$ after excitation at $t=0$ is $p(t)=\tau^{-1}e^{-\frac{t}{b\tau}}$. Here $\tau$ is the excited state lifetime (28 ns for Rb $5P_{1/2}$) and $b$ is the probability for radiative decay. If the sample is submitted to a continuous photon flux $\phi$ until $t=0$, the probability for detection of a photon at $t>0$ is:
$P(t) = \phi bp_0\eta \frac{\left( 1-p_0\right) }{\left(1-bp_0\right)}e^{-\frac{t}{\tau^{\prime}}}$ where $\tau^{\prime}=b\tau (1-bp_{0})^{-1}$ is the total fluorescence decay time and $\eta$ represents the detection efficiency for emitted photons escaping the medium \emph{relative} to that for laser photons \cite{supplemental}. As a consequence of the collisions with the pores walls, $\tau^{\prime}$ is bounded to the non-radiative decay time $\tau_{c}\equiv\tau b(1-b)^{-1}$.

The increase in number of photon absorption-emission cycles, due to radiation trapping, enhances the probability for non-radiative decay. This can be characterized by the energy conservation ratio (detected fluorescence photons)/(absorbed photons): $R=b\eta ( 1-p_0) (1-bp_0)^{-1}$. Eliminating $p_{0}$:
\begin{equation}\label{ryt}
    R=\eta \left[ 1+\frac{\tau ^{\prime }}\tau \frac{( b-1)}{b} \right]
\end{equation}

$\tau ^{\prime }$ and $R$ correspond to the experimentally accessible quantities $\theta$ and $\beta /\alpha$ respectively.

The porous medium was prepared by manually grinding Pyrex glass. The ground glass passed through a 74 $\mu m$ mesh and was retained by a 50 $\mu m$ mesh. It was introduced in a 5 $mm$ internal diameter Pyrex tube containing acetone. The same glass was used for the tube and the porous medium. After decantation, the acetone was evaporated and the tube temperature ramped to 780 $^{\circ}$C during four hours. The tube was let to slowly cool down inside the switched-off oven. After this treatment, the porous material acquires some rigidity and adheres to the containing tube. A small drop of Rb was distilled under vacuum into the tube that was then sealed. After a few days in the presence of Rb vapor, the porous medium acquires a characteristic light-blue color indicative of the penetration of Rb atoms \cite{Burchianti2006}.

\begin{figure}[htb]
\includegraphics[width=8cm, bb=55 0 580 612,clip=true]{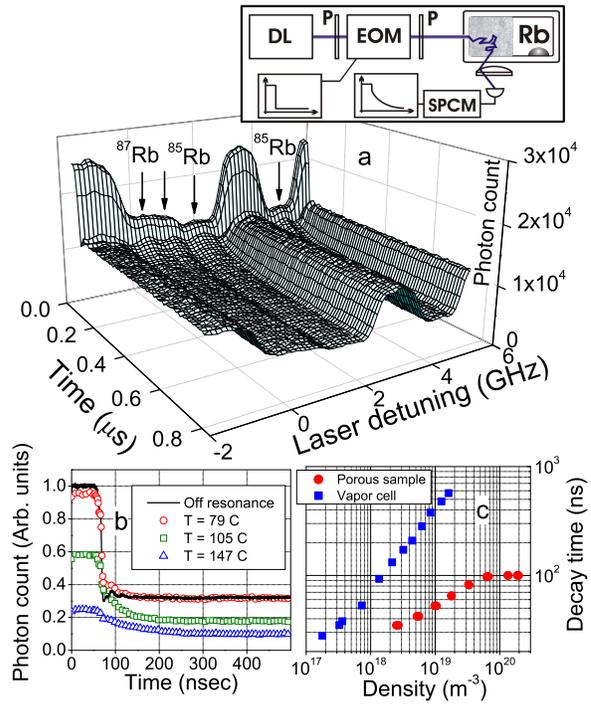}
\caption{\label{sabana}\emph{Color online} a) Photon count rate versus time and laser detuning for a sudden laser power change, $T=152$ $^{\circ}$C, arrows: transition line-centers. b) Photon count transients at several temperatures for laser frequency at the center of the $^{85}$Rb $F=3 \rightarrow F'$ line. Solid: off-resonance transient. c) Fluorescence decay-times for the $^{85}$Rb $F=2 \rightarrow F'$ line. Circles: Porous medium. Squares: vapor cell. Inset: Experimental setup. DL: Diode Laser. P: Polarizer. EOM: Electro-optical modulator. SPCM: Single photon counting module. Atomic vapor densities are estimated from temperature according to \cite{Steck:2010}.}
\end{figure}

A scheme of the experimental setup is shown in the inset of Fig. \ref{sabana}. The laser was tuned to the D1 lines of Rb  (795 nm). An electro-optic modulator (EOM), placed between crossed polarizers, was used for intensity modulation of the light. Driving the EOM with a square-wave voltage generator  switches the light intensity at the sample between two values. The switching time is of the order of $10\ ns$. The contrast of the intensity modulation is typically $50 \%$. The tube containing the porous sample was placed inside an oven with controllable temperature. The sample tube is illuminated from one end and light collected from a small region ($\sim 10^{^{-2}} mm^{2}$) of the cylindrical surface of the porous medium. A single-photon counter was used for light measurement. A saturated absorption setup allowed laser frequency calibration.

Fig. \ref{sabana}a shows a 3D plot of the transient light transmission signal in response to a stepwise variation of the laser power for different laser detunings. This plot was taken for large atomic vapor density ($T=152$ $^{\circ}$C) at which a reduction of the light transmission is clearly visible at the atomic transitions. Notice the sharp transient off resonance and the slower decay observed on resonance. Also visible in Fig. \ref{sabana}a is the distorted non-Gaussian shape of the transmission-spectrum lines to be discussed below.

A detailed plot of the transient light transmission at a fixed laser frequency (center of the $^{85}$Rb $F=2 \rightarrow F'$ Doppler broadened line) is presented in Fig. \ref{sabana}b together with an off-resonance trace. As the temperature of the cell and thus the atomic density are raised, the characteristic relaxation time increases. During the initial 20-30 $ns$ following the power switching, the transient is affected by the electronic response of the light modulation system. The tails of the slow transients observed on resonance were well fitted to exponential decay functions for the determination of $\beta$ and $\tau ^{\prime }$.

The variation of $\tau ^{\prime }$ reported here constitutes the first observation of radiation trapping inside a passive scattering medium. $\tau ^{\prime }$ appears to be bounded to about 100 $ns$ even for the largest atomic densities indicating a non-radiative decay mechanism. For comparison, we have measured fluorescence decay-times in a Rb vapor cell with no porous medium where no bound of  $\tau ^{\prime }$ was reached (Fig. \ref{sabana}c).
\begin{figure}[htb]
\includegraphics[width=8cm]{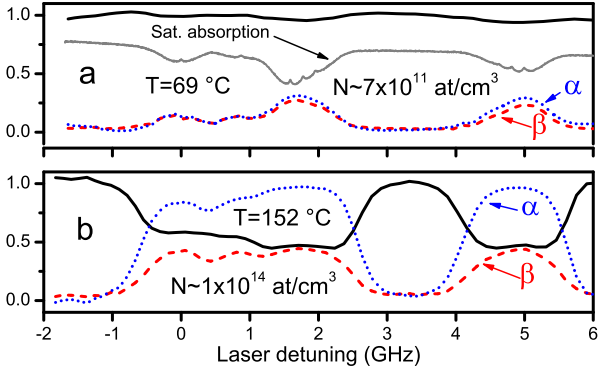}
\caption{\label{espectros} \emph{Color online} Relative transmission (solid), absorption (dotted) and fluorescence (dashed) as a function of laser detuning relative to the $^{87}$Rb $F=2 \rightarrow F'=1$ transition for two temperatures. Atomic vapor densities ($N$) are estimated from temperature according to \cite{Steck:2010}. \emph{The traces are normalized at 3.4 Ghz, small ($\lesssim 10\%$) laser power variations occur over the scan.}}
\end{figure}

The relative light transmission through the porous sample as a function of laser detuning is presented in Fig. \ref{espectros} for two different cell temperatures. Also shown are the corresponding values of $\alpha$ and $\beta$. For low atomic density ($T = 69$ $^{\circ}$C, Fig. \ref{espectros}a), the transmission presents very small structures as the laser frequency is swept across the atomic resonances. This could be erroneously interpreted as due to negligible atomic absorption. However, the corresponding value of $\alpha$ indicates a significant absorption largely compensated (to about 75\%) by the atomic fluorescence as a consequence of the ISE.

The transmission spectrum corresponding to a higher atomic density ($T = 152$ $^{\circ}$C, Fig. \ref{espectros}b), shows large structures around the atomic transitions. The corresponding values of $\alpha$ approaches total absorption. As expected for an optically thick medium, the absorption for the two Rb isotopes is not in proportion to the natural abundance. The fluorescence coefficient $\beta$ is smaller than  $\alpha$ by a factor around two. The spectra of $\alpha$ and $\beta$ show significant lineshape differences. The spectral lines are better resolved in the  $\beta$ spectrum. In consequence, the transmission spectrum presents an unusual shape with local maxima at line center.
\begin{figure}[htb]
\includegraphics[width=8cm]{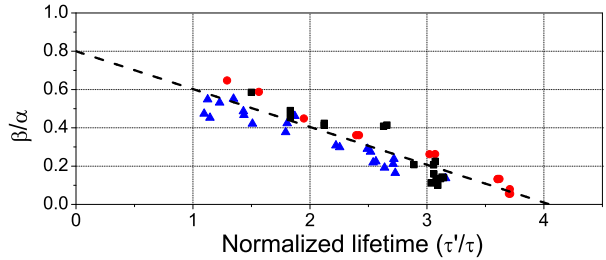}
\caption{\label{rytfig} \emph{Color online} Measured fluorescence/absorption energy ratio $\beta/\alpha$ as a function of fluorescence lifetime. Symbols refer to different choices of the sample surface imaged on the detector. Dashed: linear fit for all points.}
\end{figure}

Fig. \ref{rytfig} shows the observed dependence of $R$ on the corresponding values of $\tau ^{\prime }$. Although the atomic density spans more than two orders of magnitude, the linear dependence of Eq. \ref{ryt} is satisfactorily verified with $\eta \simeq 0.8$ and $b\simeq 0.8$.

 The typical pore size in our sample (several tenth of microns) is larger than the distance traveled by an atom at mean thermal velocity during an excited state lifetime. Under these conditions, the excited atoms colliding with the walls correspond approximately to those contained in the external pore layer of thickness $d=\tau(\frac{k_{B}T}{2\pi m})^{\frac{1}{2}}$ ($\approx 2.1\ \mu m$ at $T=400$ K) \cite{supplemental}. From this assumption, the corresponding pores surface to volume ratio is $\frac{A}{V}=\frac{1-b}{bd}$ ($\approx 1\times 10^{5}$ $m^{-1}$). For spherical pores, this corresponds to a mean pore diameter $D\simeq \frac{6bd}{1-b}$ ($D\approx 5\times 10^{-5}\ m$) consistent with the ground glass grain-size.

In summary, we have made the first spectroscopical study of an optically resonant transition in atoms contained in a porous sample. We observed that the light-path randomization by the porous medium results in a substantial integrating-sphere effect. The reduction in the fluorescence to absorption ratio, occurring for high atomic densities, appears to be the consequence of increased probability for non-radiative energy decay due to atom-wall collisions favored by radiation trapping. From the spectroscopic data, a satisfactory estimate of the pores surface to volume ration and the mean pore diameter was derived (cf. \cite{Cassidy:2011,Svensson:2010,Svensson:2012} for related work on the relation between wall collisions and pore size). Our work constitutes an initial approach to the spectroscopy of alkali atoms in materials with micrometer size porosity. Further study with different porous size and new materials is underway.

We are thankful to D. Bloch for encouraging discussions and to F. Wiotte for building the fast switching electronics. This work was supported by ANII, CSIC, PEDECIBA (Uruguay) and ECOS (France-Uruguay).


\pagebreak


\end{document}


\twocolumn[ 

\title{Rb optical resonance inside a random porous medium}

\author{S. Villalba$^{1}$, H. Failache$^{1,*}$, A. Laliotis$^{2,1}$, L. Lenci$^{1}$, S. Barreiro$^{1}$ and A. Lezama$^{1}$}

\address{
$^1$Instituto de F\'{\i}sica, Facultad de Ingenier\'{\i}a,
Universidad de la Rep\'{u}blica,\\ J. Herrera y Reissig 565, 11300
Montevideo, Uruguay\\
$^2$Laboratoire de Physique des Lasers UMR 7538 du CNRS, Universit\'{e} Paris-13, F-93430, Villetaneuse, France\\
$^*$Corresponding author: heraclio@fing.edu.uy
}
\title{Supplemental material}

 ] 

\section{\label{trapping} Radiation trapping and energy conservation factor}
\noindent In the case of a closed transition, at high atomic densities, the characteristic decay time of the atomic fluorescence is affected by the onset of photon-trapping: a photon emitted by an atom has a significant probability of being re-absorbed by another atom. Consequently, as the number of absorption and re-emission cycles grows, the characteristic decay time of the atomic fluorescence is increased. In the following we present a simplified model of the \emph{mean} evolution of the radiative decay in an atomic sample in the presence of photon trapping. More detailed treatments can be found in \cite{Andreas1998}.

Let $p_{0}$ be the probability for a photon present in the atomic medium to be absorbed by an atom [$(1-p_{0})$ is then the probability for the photon to escape de medium without absorption]. In this crude model we ignore any dependence of $p_{0}$ on time, position or photon frequency. $p_{0}$ essentially depends on the medium optical density $\Delta$.

\begin{equation}\label{probabilidad}
p_{0}\sim 1-e^{-\Delta}
\end{equation}

We consider two possible atomic de-excitation mechanisms: radiative decay with a characteristic rate $\tau^{-1}$ and non-radiative decay due to collision with the pore walls at a rate $\tau_{c}^{-1}$. The total decay rate is $\tau_{tot}^{-1}=\tau^{-1}+\tau_{c}^{-1}$ After excitation at time $t=0$ the probability density for the atom to emit a photon at time $t$ is given by

\begin{equation}\label{decaimiento}
p(t)=b\tau_{tot}^{-1}e^{-\frac{t}{\tau_{tot}}}= \tau^{-1}e^{-\frac{t}{b\tau}}
\end{equation}

In Eq. \ref{decaimiento} we have introduced the branching ratio for radiative emission:
\begin{equation}\label{branching}
b\equiv \frac{\tau_{c}}{\tau+\tau_{c}}
\end{equation}
>From Eq. \ref{decaimiento} one can calculate the probability $P_{n}(t)$ for detection at time $t$, after $n$ absorption-emission cycles, of a photon launched into the medium at $t=0$.

\begin{equation}\label{nfotones}
P_n\left( t\right)  =\eta \left( 1-p_0\right) \left( \frac{p_0}\tau
\right) ^n\frac{t^{n-1}}{\left( n-1\right) !}e^{-\frac t{b\tau} }
\end{equation}
here $\eta$ represents the detection efficiency for emitted photons escaping the medium \emph{relative} to that for laser photons that would reach the detector in the absence of atoms. $\eta$ can be understood as a figure of merit for the Integrating Sphere Effect (ISE) since $\eta=1$ would correspond to the same probability of detection for laser or fluorescence photons and $\eta=0$ would correspond to the usual absorption setup where the contribution of fluorescence photons is negligible. Also note that $\eta$ being a ratio it is insensitive to the finite detector quantum efficiency.

Summing over $n$, the total probability $P(t)$ for detection at time $t$ of a photon launched into the medium at $t=0$ is:

\begin{equation}\label{probabilidadsumada}
P\left( t\right) =\sum_{n=1}^\infty P_n\left( t\right) =\frac{p_0\eta
\left( 1-p_0\right) }\tau e^{-\frac{\left( 1-bp_0\right) t}{b\tau} }
\end{equation}

Consider now that the atomic medium is submitted to a continuous photon flux $\phi$ which is turned off at $t=0$ then the probability density for photon detection at $t>0$ is:
\begin{eqnarray}
\nonumber
  \overline{P}(t) &=& \phi \int_{-\infty }^0P\left( t-t^{\prime}\right) dt^{\prime } \\
  \overline{P}(t) &=& \phi bp_0\eta \frac{\left( 1-p_0\right) }{\left(1-bp_0\right) }e^{-\frac{ t}{\tau^{\prime}} }
\end{eqnarray}
where we have introduced the total decay time
\begin{equation}\label{tauprima}
    \tau^{\prime}=\frac{b\tau}{1-bp_{0}}
\end{equation}

It is worth noticing that the non-radiative decay mechanisms represented by $b$ introduces an upper bound for $\tau^{\prime}$ i.e. $\tau^{\prime}_{MAX} = \tau b(1-b)^{-1}\equiv \tau_{c}$.

The total probability for detecting a photon after absorption at $t=0$ is:
\begin{equation}\label{probabilidadtotal}
   \overline{p}=\int_{0}^{\infty}P(t)dt=\frac{b p_{0}\eta \left( 1-p_0\right) }{1-bp_0}
\end{equation}

We can define the energy conservation factor $R=\overline{p}/p_{0}$ that characterizes the probability for a photon initially absorbed to result in a detected fluorescence photon. From Eq. \ref{probabilidadtotal} we have:

\begin{equation}\label{compensacion}
   R=\frac{b\eta \left( 1-p_0\right) }{1-bp_0}
\end{equation}

A connection between the energy conservation factor $R$ and the fluorescence decay time $\tau^{\prime}$ can be obtained by eliminating $p_{0}$ between Eqs. \ref{tauprima} and \ref{compensacion}.
\begin{equation}\label{ryt}
    R=\eta \left[ 1+\frac{\tau ^{\prime }}\tau \frac{(b-1)}{b} \right]
\end{equation}

$\tau ^{\prime }$ and $R$ correspond to the experimentally accessible parameters $\theta$ and $\beta /\alpha$ respectively.

\section{\label{porosidad} Porosity parameters estimation}
The distribution for an atomic velocity component $v$ is given by the Gaussian Maxwell-Boltzmann distribution:
\begin{equation}\label{gaussiana}
    f(v)=(\frac{m}{2\pi k_{B}T})^{\frac{1}{2}}e^{-\frac{v^{2}}{2\sigma^{2}}}
\end{equation}
where the mean square velocity is $\sigma^{2}=\frac{k_{B}T}{m}$

The mean \emph{positive} velocity in a given direction is:
\begin{equation}\label{velocidadpositiva}
    \overline{v}_{out}=\int_{0}^{+\infty}f(v)v dv=(\frac{k_{B}T}{2\pi m})^{\frac{1}{2}}
\end{equation}

The distance traveled by an average excited atom at the mean positive velocity is $d=\overline{v}_{out} \tau$ ($d\simeq 2.1\  \mu m$ for Rb at 400 K).

\subsection{\label{supvol} Surface to volume ratio}
Assuming that the typical pore size is large compared to $d$, the total atomic flux at the pores walls is:

\begin{equation}\label{flujoatomico}
    J=\rho  \overline{v}_{out}
\end{equation}
where $\rho=\frac{N}{V}$ is the atomic density ($N$ is the total number of atoms and $V$ the internal pore volume).
The wall collision rate per atom is then:
\begin{equation}\label{tasacontrlapared}
    \tau_{c}^{-1}=\frac{(1-b)}{b\tau} = \overline{v}_{out} \frac{A}{V}
\end{equation}
or
\begin{equation}\label{areasobrevolumen}
    \frac{A}{V}=\frac{(1-b)}{bd}
\end{equation}
where $A$ is the total area of the pores surface.

\subsection{\label{diametro} Mean pore size estimation}
Considering a spherical pore, the surface to volume ratio given by \ref{areasobrevolumen} corresponds to a pore diameter:
\begin{equation}\label{D}
    D\simeq \frac{6bd}{1-b}
\end{equation}

For the experimentally determined value of $b=0.8$, we obtain $D=5\times 10^{-5}\ m$ in consistency with the approximation $D\gg d$.

